\documentclass[aps,prl,twocolumn,showpacs,superscriptaddress,groupaddress]{revtex4}
\usepackage{graphicx}  
\usepackage{dcolumn}   
\usepackage{bm}        
\usepackage{amssymb}   
\usepackage{color}

\newcommand{\comment}[1]{}

\begin{document}

\title{Stabilization of quantum metastable states by dissipation}

\author{D. Valenti\footnote{e-mail address: davide.valenti@unipa.it}}
\affiliation{Dipartimento di Fisica e Chimica, Group of Interdisciplinary Physics, Universit\`{a} di Palermo and CNISM,
Unit\`{a} di Palermo, Viale delle Scienze, Edificio 18, I-90128 Palermo, Italy}
\author{L. Magazz\`{u}\footnote{e-mail address: luca.magazzu@unipa.it}}
\affiliation{Dipartimento di Fisica e Chimica,
Group of Interdisciplinary Physics,
Universit\`{a} di Palermo and CNISM, Unit\`{a} di Palermo,
Viale delle Scienze, Edificio 18, I-90128 Palermo, Italy}
\affiliation{Radiophysics Department, Lobachevsky State University of Nizhni Novgorod, 23 Gagarin Avenue, Nizhni Novgorod 603950, Russia}
\author{P. Caldara}
\affiliation{Dipartimento di Fisica e Chimica,
Group of Interdisciplinary Physics,
Universit\`{a} di Palermo and CNISM, Unit\`{a} di Palermo,
Viale delle Scienze, Edificio 18, I-90128 Palermo, Italy}
\author{B. Spagnolo \footnote{e-mail address: bernardo.spagnolo@unipa.it}}
\affiliation{Dipartimento di Fisica e Chimica,
Group of Interdisciplinary Physics,
Universit\`{a} di Palermo and CNISM, Unit\`{a} di Palermo,
Viale delle Scienze, Edificio 18, I-90128 Palermo, Italy}

\date{\today}

\begin{abstract}

Normally, quantum fluctuations enhance the escape from metastable states in the presence of dissipation. 
Here we show that dissipation can enhance the stability of a quantum metastable system, consisting of a 
particle moving in a strongly asymmetric double well potential, interacting with a thermal bath. We find that 
the escape time from the metastable state has a nonmonotonic behavior versus the system-bath coupling
and the temperature, producing a stabilizing effect. 

\end{abstract}

\pacs{03.65.Aa, 03.65.Xp, 03.65.Yz, 03.65.Db}

\keywords{Caldeira-Leggett model; metastable potential; discrete variable representation; noise enhanced stability}
\maketitle

\emph{Introduction}. $-$ Recently, the role of dissipation on the dynamics of quantum systems has been the subject of 
renewed interest~\cite{Kas13}. 

The presence of a dissipative environment indeed influences significantly the escape from a quantum metastable state.
This is a general problem, of interest in many areas of physics, whenever a sudden change in the state of a system 
occurs on timescales small with respect to the typical times of the systemÕs dynamics.

The archetypical model describing the escape process is that of a particle subject to a cubic or asymmetric bistable 
potential and linearly coupled to a heat bath of quantum harmonic oscillators~\cite{Cal81,Gri98,Wei08}. In such a 
system the decay from the metastable state occurs on timescales that depend on the friction and temperature. 
Various physical systems such as magnetization in solid state systems~\cite{Gatt03}, proton transfer in chemical 
reactions~\cite{Nau99} and  superconducting devices~\cite{Nak99} can be described within this framework.

Calculations of the decay rates, using a cubic potential, have been performed in Refs.~\cite{Aff81,Grab87} 
using functional integral techniques. In Ref.~\cite{Grab87}, starting with the particle at the bottom of the 
metastable well, it has been shown that  the decay rate decreases monotonically as  the damping increases 
and grows with the bath temperature\comment{with a non zero value in absence of interaction with the bath}. 
Similarly, by using a master equation technique, a monotonic increase of the escape rate, with respect to
the temperature, is found in Ref.~\cite{Sarg07} for a Gaussian wave packet initially in the metastable well of 
a biased quartic potential.

Stabilization of a quantum metastable state by an external time-periodic driving, in absence of environment, 
was obtained in Ref.~\cite{Choo05}. Moreover, suppression of activated escape from a metastable state by 
increasing the temperature was found in a time-periodically driven quantum dissipative system~\cite{Shi13}.

Common wisdom is that environmental fluctuations always enhance the escape from a quantum metastable 
state. A critical issue and of great importance is: can the dissipation enhance the stability of a quantum 
metastable state?

To answer this question we follow the time evolution of the populations of spatially localized states in a strongly 
asymmetric bistable system, starting from a nonequilibrium initial condition.  This choice allows us to observe how, 
increasing the damping, the relaxation process towards the stable well goes from a population transfer in which the 
metastable well is temporarily populated, to a mechanism of direct transfer to the stable state. This stabilization effect 
is related to that due to the suppression of tunneling by dissipation in quantum regime~\cite{Cal81,Grab87}. As a result 
we find that dissipation can enhance the stability of the quantum metastable state. Indeed, we observe 
that the escape dynamics is characterized by a nonmonotonic behavior, with a maximum, as a function of the 
damping strength: there is an optimal value of the damping strength which maximizes the escape time, producing 
a \emph{stabilizing} effect in the quantum system. This result, which resembles the phenomenon known, in the 
classical context, as noise enhanced stability (NES) of metastable states~\cite{Man96,Agu01}, sheds new
light on the role of the environmental fluctuations in stabilizing quantum metastable systems. We also find that the
behavior of the escape time versus the temperature is nonmonotonic with a minimum. Therefore as the temperature 
increases, an enhancement of the escape time is observed, increasing the stability of the quantum metastable state.

\begin{figure}[htbp]
\begin{center}
\includegraphics[width=7cm,angle=0]{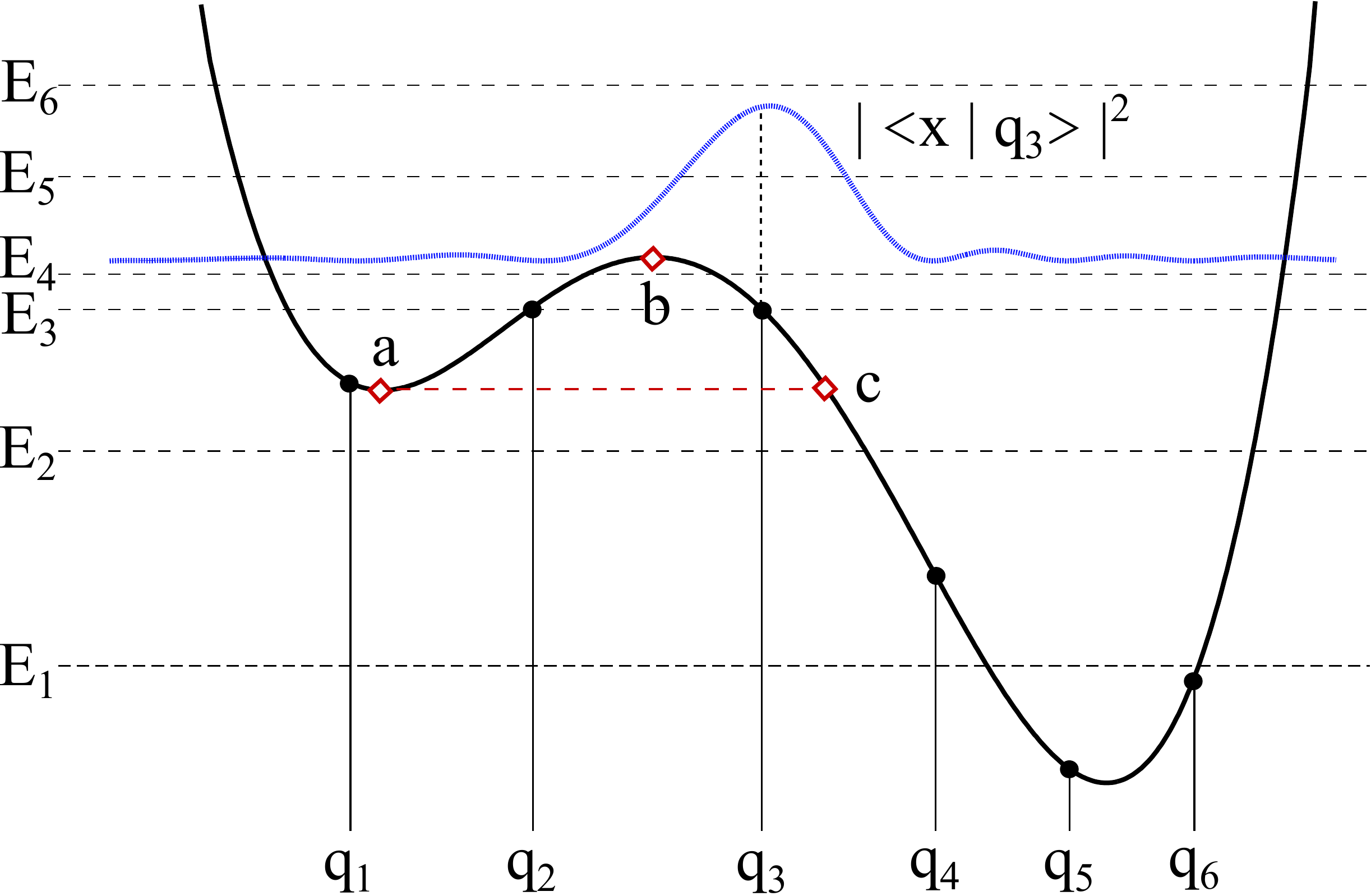}
\end{center} \caption{(Color online) Potential profile $V(q)$ (Eq.~(\ref{bistable
potential})) for $\Delta U = 1.4\hbar\omega_0$ and $\epsilon = 0.27
\sqrt{M\hbar\omega_0^3}$. Horizontal lines: the first $6$
energy levels. Vertical lines: the position eigenvalues in the DVR. The dashed curve
is the initial probability distribution $|\Psi(x,0)|^{2}$. For the tunneling splitting we have
$\Delta E_{4,3}= E_{4}-E_{3} = 0.2\hbar\omega_{0}$  while $E_{2}-E_{1} = 0.985\hbar\omega_{0}$. }
\label{fig1}
\end{figure}

\emph{Model}. $-$ We consider a quantum \emph{particle} of effective mass $M$ in a double well potential 
(see Fig.~\ref{fig1}). The system's bare Hamiltonian is $\hat{H}_0=\hat{p}^{2}/2M+V (\hat{q})$, where
\begin{equation}
V (\hat{q})=\frac{M^2\omega^4_0}{64\Delta
U}\hat{q}^4-\frac{M\omega^2_0}{4}\hat{q}^2-\hat{q}\epsilon.
\label{bistable potential}
\end{equation}
\noindent Here $\omega_0$ is the natural oscillation frequency around the minima, $\Delta U$ is
the barrier height and $\epsilon$ the asymmetry parameter. In this work $\epsilon$ is large enough to mimic
the cubic potential, which is the archetypal model potential for metastable systems.

The environment is a thermal bath of $N$ independent harmonic oscillators with position coordinates 
$\hat{x}_j$ and coupled with the particle through the  linear interaction  term 
$\sum_{j=1}^N c_j\hat{x}_j\hat{q}$.

In our study it is assumed that, in the continuum limit $N\rightarrow\infty$, the bath has the Ohmic 
spectral density $J(\omega)=M\gamma\omega$  with a cut-off at a frequency much larger than any 
other involved in the model. The damping constant $\gamma$ is a measure of the overall particle-bath 
coupling strength and has the same role as the classical damping in the quantum Langevin equation
associated to the present problem~\cite{Wei08}.

In the quantum regime, given the particle's initial preparation, the bath temperature is such that the
dynamics is practically confined among the first $6$ levels of the potential shown in Fig.~\ref{fig1}.
In this reduced Hilbert space, performing a suitable transformation, we pass to the discrete
variable representation (DVR) \cite{Har65}, in which the particle's reduced dynamics is described in
terms of the \emph{localized} basis of the position eigenstates $\{|q_1\rangle,\dots,|q_6\rangle\}$, where
$\hat{q}|q_i\rangle = q_i |q_i\rangle$.

\emph{Analytical method}. $-$ We assume a factorized initial condition, with the bath in the thermal state
$\rho_B(0)=e^{-\beta\hat{H}_B}/Z_B$. The particle's reduced density operator in the DVR is given by
\begin{equation}\label{density-operator}
\rho_{\mu\nu}(t)=\sum_{\alpha,\beta=1}^6
K(q_{\mu},q_{\nu},t;q_{\alpha},q_{\beta},0)\rho_{\alpha\beta} (0),
\end{equation}
\noindent where $\rho_{\mu\nu}(t)=\langle q_{\mu}|\rho(t)|q_{\nu}\rangle$ ($\mu,\nu=1,\dots,6$) and
$K(q_{\mu},q_{\nu},t;q_{\alpha},q_{\beta},0)$ is given by the double dissipative path integral
\begin{equation}\label{K}
  \int_{q_{\alpha}}^{q_{\mu}}\mathcal{D}q(t)
\int_{q_{\beta}}^{q_{\nu}}\mathcal{D}^*q'(t)\mathcal{A}[q]
\mathcal{A^*}[q'] \mathcal{F}_{FV}[q,q'].
\end{equation}
\noindent Here $\mathcal{A}[q]$ is the amplitude associated with the path $q(t)$ of the bare system.
In the DVR a path is a step-like function with transitions among the positions $q_i$. The effect exerted by
the bath on the quantum mechanical amplitude associated to a path ($q(t), q\prime(t)$) is condensed in the
Feynman-Vernon (FV) influence functional $\mathcal{F}_{FV}[q,q']$~\cite{Fey63}.

This approach is non-perturbative in the system-bath coupling, and is thus suited for dealing with the
strong coupling regime. Nevertheless, the FV influence functional makes the path integral intractable as it 
introduces time nonlocal \emph{interactions} between the paths $q(t)$ and $q\prime(t)$, through the bath 
correlation function $Q(t)$.

The nonlocal part of the interactions cancels out in the limit in which the bath correlation function $Q(t)$ is linear 
in $t$, i.e. in the long time limit $t\gg\hbar/k_{B}T$ (see section IV and appendix B in Ref~\cite{Tho01}). If 
the temperature is sufficiently high, $Q(t)$ can be taken in the linearized form at all times, which amounts to perform 
the so called \emph{generalized non-interacting cluster approximation} (gNICA~\cite{Tho01}, the multi-level version 
of the NIBA for the spin-boson model~\cite{Cal81,Wei08}). By comparing the transition probabilities per unit time among 
the $|q_{i}\rangle$'s with $k_{B}T/\hbar$, we obtain the limit $T\gtrsim 0.1 \hbar \omega_0 / k_{B}$ as a rough estimate 
for the validity of the gNICA for our system.  

Within the gNICA the double path integral of Eq.~(\ref{K}) assumes a factorized form in the Laplace space, allowing 
for the derivation of a generalized master equation (GME). If $\rho(0)$ is diagonal in the position representation, the 
GME reads~\cite{Tho01} 
\begin{equation}
\dot{\rho}_{\mu\mu}(t)=\sum_{\nu=1}^{6}\int_{0}^t dt'\mathcal{H}_{\mu\nu}(t-t'){\rho}_{\nu\nu}(t').
\label{GME}
\end{equation}
The elements of $\mathcal{H}$ are taken to the second order in the transition amplitudes per unit time 
$\Delta_{ij}=\langle q_i|\hat{H}_0| q_j\rangle/\hbar$  and at all orders in the system-bath coupling. 
They display a cutoff of the type $\exp(-\gamma t \times \text{const.} )$. At strong damping (as in our case, see below), 
in the short time interval in which $\mathcal{H}_{\mu\nu}$ is substantially different from zero, $\rho_{\nu\nu}(t)$ is 
practically constant. This allows us to write the following rate equation as the Markov approximated version of Eq.~(\ref{GME})
\begin{equation}
\dot{\rho}_{\mu\mu}(t)=\sum_{\nu=1}^6\Gamma_{\mu\nu}\rho_{\nu\nu}(t),
\label{ME}
\end{equation}
where $\Gamma_{\mu\nu}=\int_{0}^{\infty}d\tau \mathcal{H}_{\mu\nu}(\tau)$.
\comment{The solution of Eq.~(\ref{ME}) has the form
\begin{equation}
\rho_{\mu\mu}(t)=\sum_{\nu,\alpha=1}^6 c_{\mu\alpha}.
e^{\Lambda_\nu t}\rho_{\alpha\alpha}(0)
\label{solution}
\end{equation}
\noindent where $\Lambda_{\nu}$ are the eigenvalues of $\Gamma$.}

The smallest, in absolute value, of the nonzero eigenvalues of the rate matrix $\Gamma$ determines the largest time-scale of 
the dynamics, that is the quantum relaxation time $\tau_{\text{relax}}$~\cite{Tho01}.

\emph{Escape time}. $-$ In the following we focus on the particle's transient
dynamics, as given by the solution of Eq.~(\ref{ME}) with the nonequilibrium initial condition
\begin{equation}\label{init-cond}
 \rho(0)=|q_3\rangle\langle q_3|
 \end{equation}
i.e. with the particle's probability density initially peaked on the right of the potential
barrier, in the interval ($q_b, q_c$) (see Fig.~\ref{fig1}). This may be experimentally attained by preparing the particle in the 
ground state of an appropriate harmonic well centered at the desired position, and then
releasing the harmonic potential~\cite{Chi12}.

Before giving the definition of escape time in the present context, we define the population of the lower (right side) well
 as the cumulative population of the three DVR states from $|q_4\rangle$ to $|q_6\rangle$, that is
\begin{equation}
P_{\text{right}}(t)=\sum_{\mu=4}^6 \rho_{\mu\mu}(t),
\label{right population}
\end{equation}
which is a discretized version of the probability of penetration of the wave packet through the
barrier~\cite{Sarg07}.
During the transient dynamics the populations
of the metastable states ($|q_1\rangle$ and
$|q_2\rangle$) reach a maximum.
Afterwards, by tunneling through the potential barrier, the
population of the metastable well decays, finally settling down
to a stationary value dependent on the temperature.

We consider large asymmetry of the potential, low temperatures with respect to the barrier height,
and damping regimes ranging from moderate to strong ($\gamma\gtrsim \omega_0$).
Given the above conditions, the relaxation occurs in the incoherent regime, with no oscillations in
the populations. As a consequence we may consider the particle irreversibly escaped from
the metastable state once $P_{\text{right}}(t)$ has reached a certain threshold value that we set to
$P_{\text{right}}(\tau) = 0.95$.

\begin{figure}[htbp]
\begin{center}
\includegraphics[width=9cm,angle=0]{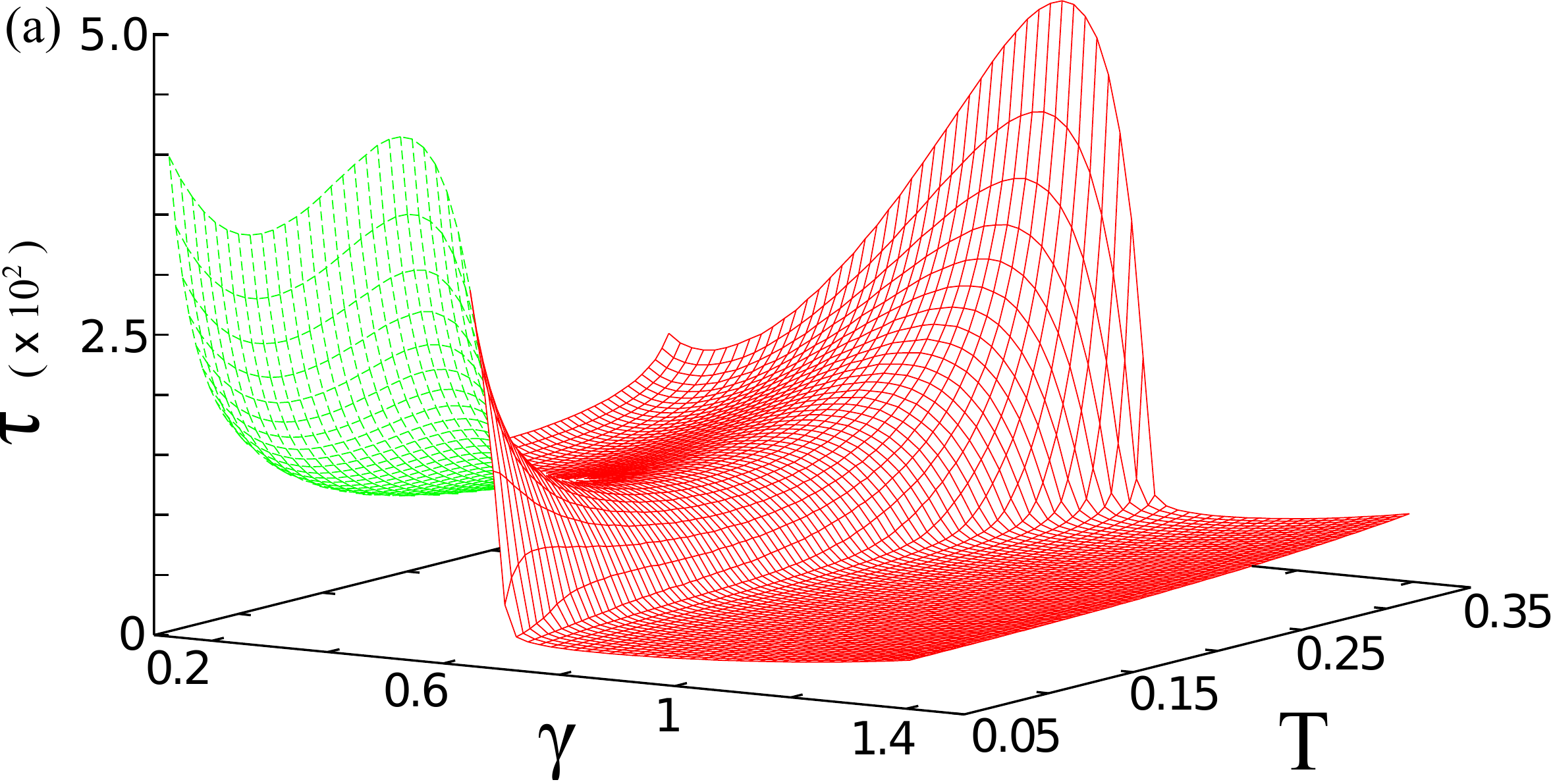} \\
\includegraphics[width=9cm,angle=0]{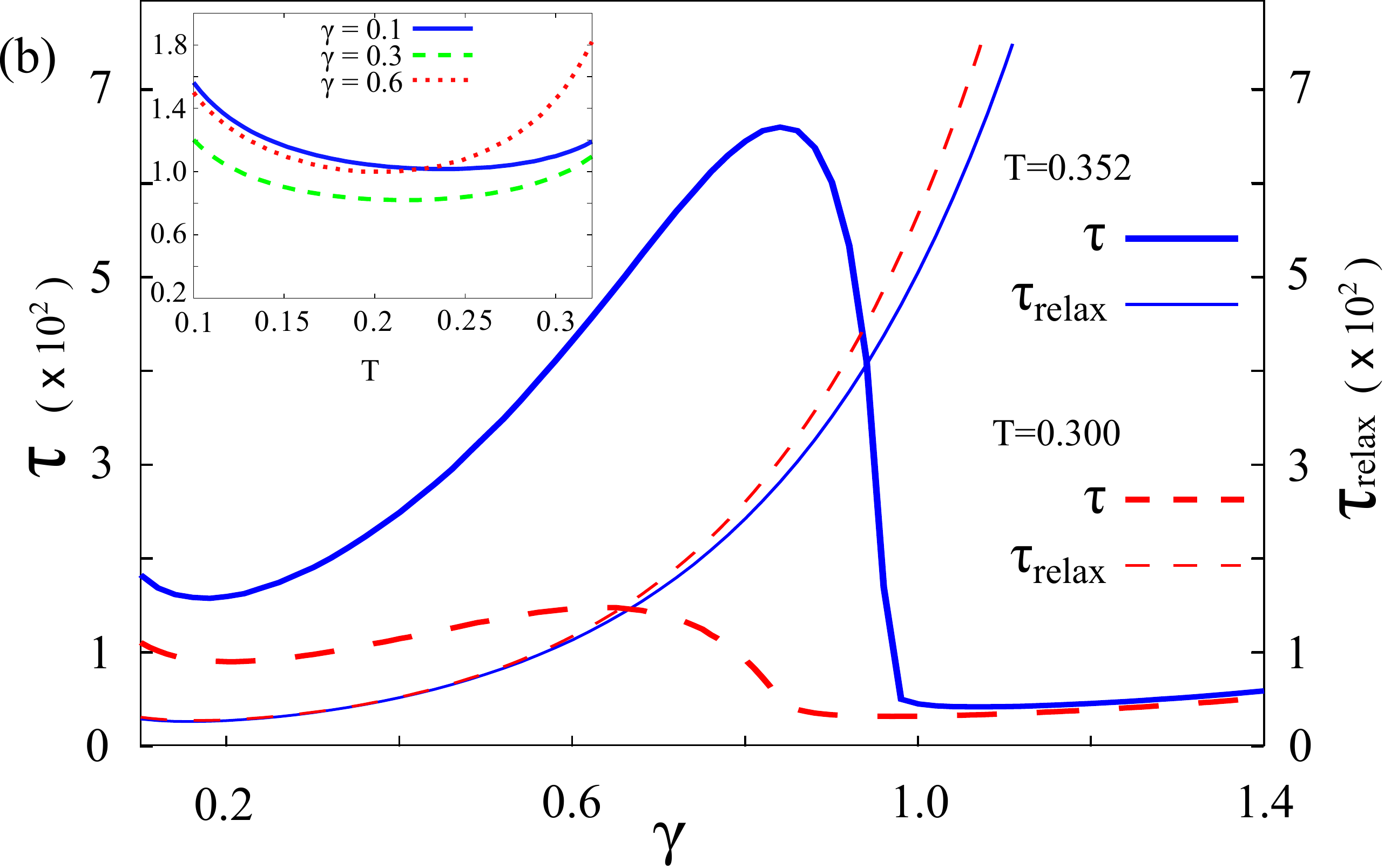}
\caption{(Color online) Escape time $\tau$, in units of $\omega_0^{-1}$, for the initial condition $ \rho(0)=|q_3\rangle\langle q_3|$
(see Eq.~(\ref{init-cond})). (a) $\tau$ and $\tau_{\text{relax}}$ as a function of both damping $\gamma$ and 
bath's temperature $T$. (b) $\tau$ as a function of $\gamma$ for different temperatures. Inset - escape
time vs temperature at fixed values of $\gamma$. The variables $\gamma$ and $T$ are given in units of 
$\omega_0$ and $\hbar \omega_0/k_B$, respectively.}
\label{fig2}
\end{center}
\end{figure}

\emph{Results}. $-$ 
We observe a nonmonotonic behavior of $\tau$ with respect to both $\gamma$ and $T$ . 
 In Fig.~\ref{fig2} it is shown the presence of a peak in $\tau$ vs $\gamma$, whose height and position depend on the temperature. 

A comparison between $\tau$ and $\tau_{\text{relax}}$ versus $\gamma$ indicates that the two  quantities exhibit roughly 
the same behavior until the peak in $\tau$ is reached (see Fig.~\ref{fig2}b). At higher $\gamma$, while $\tau_{\text{relax}}$  continues to 
increase monotonically, $\tau$ has a sudden fall off at a critical value $\gamma_c$, dependent on the temperature 
(for example $\gamma_c \simeq 0.98$ at $T = 0.352$). 
 
This critical value corresponds to a dynamical regime in which the population transfer from the initial state to the states 
of the metastable well is inhibited and there is a direct transfer to the states of the lower right well. In this regime the probability 
to find the particle in the metastable state is practically zero throughout the entire dynamics. Indeed, while $\tau_{\text{relax}}$ 
is the time needed for the system to reach the equilibrium in the double well potential, the escape time is a relevant quantity 
for the transient dynamics, involving the crossing of the potential barrier and the depletion of the metastable well. Therefore, 
our analysis applies to the general problem of escape from a metastable well, starting from a nonequilibrium condition.

The nonmonotonic behavior of $\tau$ vs $\gamma$ can be interpreted as the quantum counterpart of the 
NES phenomenon observed in classical systems, and may be called quantum noise enhanced stability (QNES).

Another interesting feature is the presence of a slow monotonic increase of $\tau$ 
for $\gamma > \gamma_c$, which leads to the quantum Zeno effect~\cite{Pas05}.

The behavior of $\tau$ vs the temperature is characterized by a minimum as $k_B T$ approaches the tunneling splitting 
$\Delta E_{4,3} = E_4-E_3$ (see Fig.~\ref{fig1}). This is the signature of the \emph{thermally activated tunneling}, an 
experimentally well established  phenomenon~\cite{Frie96}. This is better shown in the inset of Fig.~\ref{fig2}b.

\begin{figure}[htbp]
\begin{center}
\includegraphics[width=7cm,angle=0]{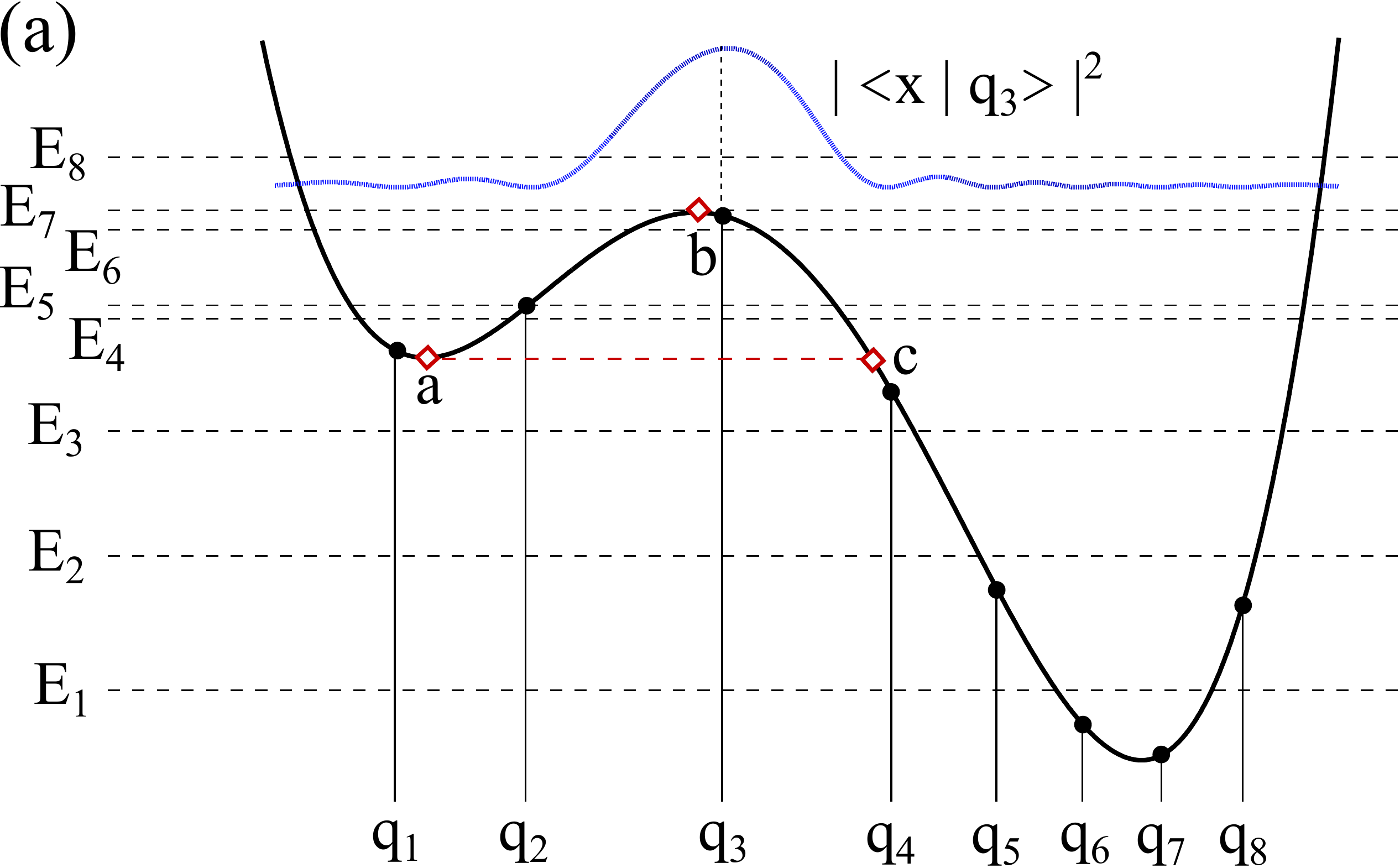}
\vskip0.3cm
\includegraphics[width=8cm,angle=0]{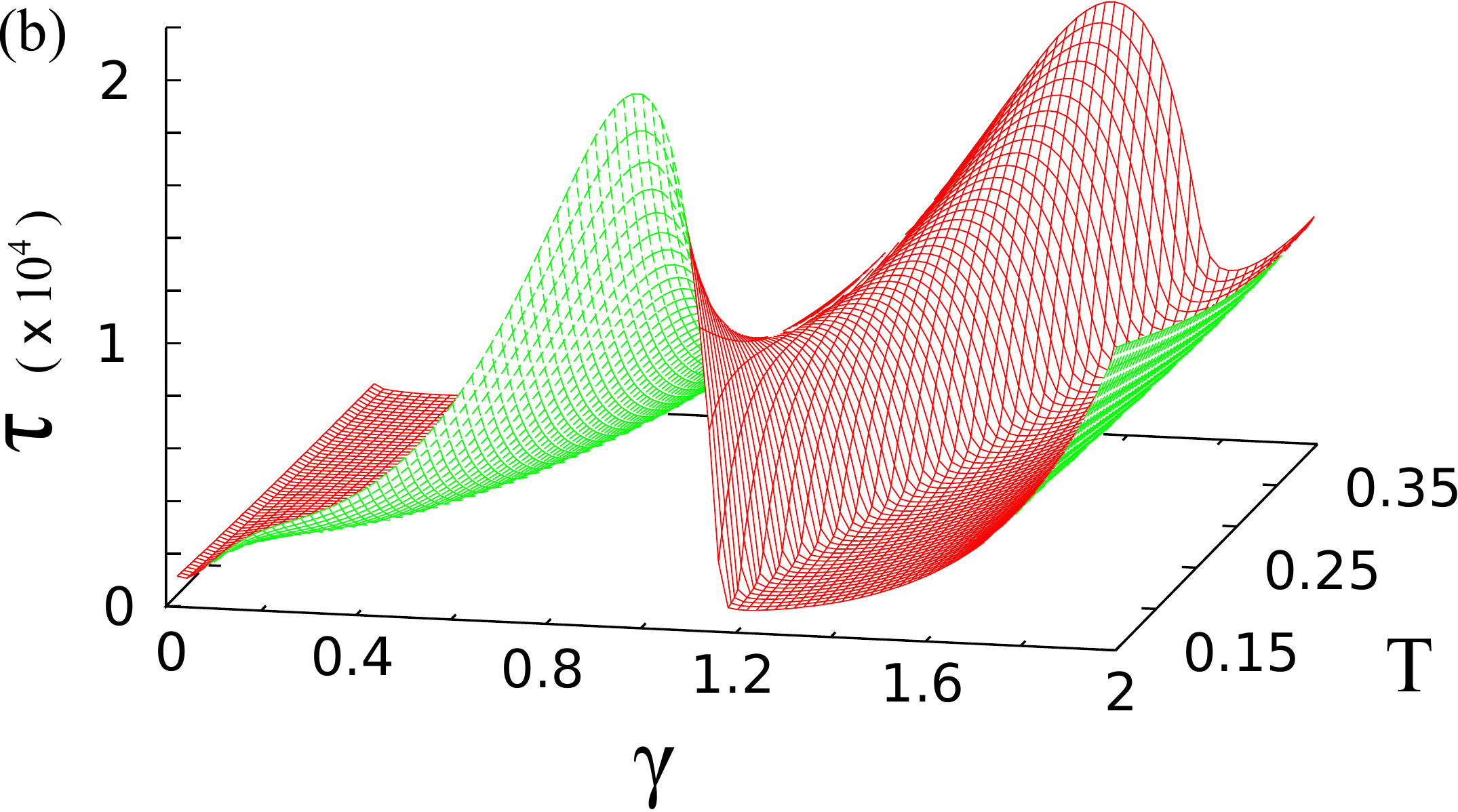}
\end{center} \caption{(Color online) (a) Potential profile $V(q)$ (see Eq.~(\ref{bistable
potential})) for $\Delta U = 2.5\hbar\omega_0$ and $\epsilon = 0.35
\sqrt{M\hbar\omega_0^3}$. Horizontal lines: the first $8$
energy levels. Vertical lines: position eigenvalues in the DVR. The dashed
curve is the initial probability distribution $|\Psi(x,0)|^{2}$. Here $\Delta E_{7,6} = 0.14\hbar\omega_{0}$,
$\Delta E_{6,5} = 0.58\hbar\omega_{0}$, $\Delta E_{5,4} = 0.1\hbar\omega_{0}$. (b) Escape time $\tau$, in units
of $\omega_0^{-1}$ for the initial condition shown in panel (a).}
\label{fig3}
\end{figure}

Finally we wish to point out that our results are robust against the variation of the potential asymmetry, threshold value
and initial conditions (initial DVR states within the interval ($q_b, q_c$), see Figs.~\ref{fig1} and ~\ref{fig3}).

To exemplify this robustness, we give here the results for another potential profile and a
different initial condition (see Fig.~\ref{fig3}).

The definition of $\tau$ is the same as for the previous case i.e. $P_{\text{right}}(\tau)=0.95$. In this
situation however, due to the different number of energy levels considered, the right well population is defined as
$P_{\text{right}}(t)=\sum_{\mu=5}^8 \rho_{\mu\mu}(t)$.

The escape time displays qualitatively the same features as for the first configuration, even if now the
initial wave packet is centered close to the top of the potential barrier. In particular $\tau$ has a
nonmonotonic behavior as a function of both $\gamma$ and $T$. The minimum of $\tau$ vs $T$ is at
$T \simeq 0.27/k_B$, which is the average value of the three tunneling splittings $\Delta E_{7,6},
\Delta E_{6,5}$, and $\Delta E_{5,4}$ (see Fig.~\ref{fig3}a). Moreover, for $\gamma > \gamma_c$,
we observe a monotonic increasing behavior of $\tau$ leading to the quantum Zeno effect.

\emph{Summary}. $-$ We have found that in nonequilibrium dynamics the escape time from a quantum
metastable state exhibits a nonmonotonic behavior as a function of both the damping, with a maximum (QNES),
and the temperature, with a minimum at the resonance with the tunneling splitting (see Figs.~\ref{fig1}, and~\ref{fig3}).

We observe stabilization of the quantum metastable state due to the dissipation and its interplay with the temperature,
in the moderate to strong damping regime. Moreover, a suppression of the activated escape is obtained by increasing
the temperature. The stabilization phenomenon associated to our model is within the reach of existing experimental
technologies such as superconducting qubits~\cite{Chi12} and optical trapping~\cite{Kor02}.

The present model could be used to control the stability of a trapped particle in atomic optics, precision spectroscopy
and optical communication.

This work was partially supported by MIUR through Grant. No. PON$02\_00355\_3391233$,
ÒTecnologie per l'ENERGia e l'Efficienza energETICa - ENERGETICÓ.

\end{document}